\documentclass[journal]{IEEEtran}
\usepackage{graphicx}
\usepackage{cite}
\usepackage{amsmath}
\usepackage{amsfonts}
\usepackage{amssymb}
\usepackage{subfigure}
\usepackage{psfrag}

\newcommand{\bm}{\mathbf}
\newcommand{\be}{\begin{equation}}
\newcommand{\ee}{\end{equation}}
\newcommand{\bea}{\begin{eqnarray}}
\newcommand{\eea}{\end{eqnarray}}
\newcommand{\x}{{\bm x}}
\newcommand{\y}{{\bm y}}
\newcommand{\p}{{\bm p}}
\newcommand{\g}{{\bm g}}
\newcommand{\q}{{\bm q}}
\newcommand{\ba}{{\bm a}}
\newcommand{\br}{{\bm r}}
\newcommand{\e}{{\bm e}}
\newcommand{\dd}{{\bm d}}
\newcommand{\bA}{{\bm A}}

\newcommand{\bF}{{\bf F}}
\newcommand{\bD}{{\bf D}}
\newcommand{\bC}{{\bf C}}

\newcommand{\bH}{{\bf H}}

\newcommand{\bg}{{\bf g}}

\newcommand{\bh}{{\bf h}}
\newcommand{\bv}{{\bf v}}

\newcommand{\bd}{{\bf d}}
\newcommand{\bs}{{\bf s}}
\newcommand{\bzero}{{\bf 0}}

\newcommand{\I}{{\bm I }}
\newcommand{\BDFT}{{\boldsymbol{\mathcal F}}}
\newcommand{\BG}{{\boldsymbol{\mathcal G}}}
\newcommand{\BE}{{\boldsymbol{\mathcal E}}}
\newcommand{\BH}{{\boldsymbol{\mathcal H}}}
\newcommand{\BD}{{\boldsymbol{\mathcal D}}}

\newcommand{\bGamma}{\mbox{\boldmath$\Gamma$}}
\newcommand{\bOmega}{\mbox{\boldmath$\Omega$}}
\newcommand{\bPsi}{\mbox{\boldmath$\Psi$}}
\newcommand{\bpsi}{\mbox{\boldmath$\psi$}}
\newcommand{\bgamma}{\mbox{\boldmath$\gamma$}}

\newcommand{\bPhi}{\mbox{\boldmath{$\Phi$}}}

\title{Low Complexity Transceiver Design for GFDM}

\author{\normalsize Arman Farhang, Nicola Marchetti and Linda E. Doyle  
\\CTVR / The Telecommunications Research Centre, Trinity College Dublin, Ireland, \\
Email: \{farhanga, marchetn, ledoyle\}@tcd.ie\vspace{-4.5 mm}}

\begin{document}

\maketitle

\begin{abstract}
Due to its attractive properties, generalized frequency division multiplexing (GFDM) is recently being discussed as a candidate waveform for the fifth generation of wireless communication systems (5G). GFDM is introduced as a generalized form of the widely used orthogonal frequency division multiplexing (OFDM) modulation scheme and since it uses only one cyclic prefix (CP) for a group of symbols rather than a CP per symbol, it is more bandwidth efficient than OFDM. In this paper, we propose novel transceiver structures for GFDM by taking advantage of the particular structure in the modulation matrix. Our proposed transmitter is based on modulation matrix sparsification through application of fast Fourier transform (FFT) to reduce the implementation complexity. A unified receiver structure for matched filter (MF), zero forcing (ZF) and minimum mean square error (MMSE) receivers is also derived. The proposed receiver techniques harness the special block circulant property of the matrices involved in the demodulation stage to reduce the computational cost of the system implementation. We have derived the closed forms for the ZF and MMSE receiver filters. Additionally, our algorithms do not incur any performance loss as they maintain the optimal performance. The computational costs of our proposed techniques are analyzed in detail and are compared with the existing solutions that are known to have the lowest complexity. It is shown that through application of our transceiver structure a substantial amount of computational complexity reduction can be achieved.
\end{abstract}

\section{Introduction}\label{sec:Intro}
\IEEEPARstart{O}{FDM} has been the technology of choice in wired and wireless systems for years, \cite{Starr1999,Li2006,Farhang2010}. The advent of the fifth generation of wireless communication systems (5G) and the associated focus on a wide range of applications from those involving bursty machine-to-machine (M2M) like traffic to media-rich high bandwidth applications has led to a requirement for new signaling techniques with better time and frequency containment than that of OFDM. Hence, a plethora of waveforms are coming under the microscope for analysis and investigation.

The limitations of OFDM are well documented. OFDM suffers from large out-of-band emissions which not only have interference implications but it also can reduce the potential for exploiting non-contiguous spectrum chunks through such techniques as carrier aggregation. For future high bandwidth applications this can be a major drawback. OFDM also has high sensitivity to synchronization errors especially carrier frequency offset (CFO). As a case in point, in multiuser uplink scenarios where OFDMA is utilized, in order to avoid the large amount of interference caused by multiple CFOs as well as timing offsets, stringent synchronization is required which in turn imposes a great amount of overhead to the network. This overhead is not acceptable for lightweight M2M applications for example. The presence of multiple Doppler shifts and propagation delays in the received uplink signal at the base station (BS) results in some residual synchronization errors and hence multiuser interference (MUI), \cite{Moreli2007}. The MUI problem can be tackled with a range of different solutions that are proposed in \cite{Lee2012,AF2013,AFWCNC2014}. However, these lead to an increased receiver computational complexity. Thus, one of the main advantages of OFDM, i.e., its low complexity, is lost. The challenge therefore is to provide waveforms with more relaxed synchronization requirements and more localized signals in time and frequency to suit future 5G applications, without the penalty of a more complex transceiver.

There are many suggestions on the table as candidate waveforms \cite{GFDM,CB-FMT,COQAM,FC-FB,FBMCMassive2014}. In general, all of these signaling methods can be considered as filter bank multicarrier (FBMC) systems. They can be broadly broken into two categories, those with linear pulse shaping \cite{FC-FB,FBMCMassive2014} and those with circular pulse shaping, \cite{GFDM,CB-FMT,COQAM}. The former signals with linear pulse shaping have attractive spectral properties, \cite{Farhang2011}. In addition, these systems are resilient to the timing as well as frequency errors. However, the ramp-up and ramp-down of their signal which are due to the transient interval of the prototype filter result in additional latency issues. In contrast, FBMC systems with circular pulse shaping remove the prototype filter transients thanks to their so called tail biting property, \cite{GFDM}. The waveform of interest in this paper is known as generalized frequency division multiplexing (GFDM) and it can be categorized as an FBMC system with circular pulse shaping. The focus of the paper, more specifically, is on the design of low complexity transceivers for GFDM.

GFDM has attractive properties and as a result has recently received a great deal of attention. One of the main attractions of GFDM is that it is a generalized form of OFDM which preserves most of the advantageous properties of OFDM while addressing its limitations. As Datta and Fettweis have pointed out in \cite{GFDM_CC}, GFDM can provide a very low out-of-band radiation which removes the limitations of OFDM for carrier aggregation. It is also more bandwidth efficient than OFDM since it uses only one cyclic prefix (CP) for a group of symbols in its block rather than a CP per symbol as is the case in OFDM. Through circular filtering, GFDM removes the prototype filter transient intervals and hence the latency. Additionally, its special block structure makes it an attractive choice for the low latency applications like IoT and M2M, \cite{GFDM_5GTC}. Filtering the subcarriers using a well-designed prototype filter limits the intercarrier interference (ICI) only to adjacent subcarriers which reduces the amount of leakage between subcarriers and increases the resiliency of the system to CFO as well as narrow band interference. In other words, GFDM has robustness to synchronization errors. As Michailow et al report in \cite{GFDM_5GTC}, GFDM is also a good match for multiple input multiple output (MIMO) systems.

The advantages of GFDM come at the expense of an increased bit error rate (BER) compared with OFDM. This degradation is due to the fact that GFDM is a non-orthogonal waveform. Consequently, non-orthogonality of the neighboring subcarriers and time slots results in self-interference. To tackle this self-interference, \textit{matched filter} (MF), \textit{zero forcing} (ZF) or \textit{minimum mean square error} (MMSE) receivers can be derived \cite{GFDM_ber}. Since, the MF receiver cannot completely remove the ICI, ZF receiver can be utilized. However, due to its noise enhancement problem, ZF receiver incurs some BER performance loss. Thus, the MMSE approach can be chosen to reduce the noise enhancement effect and maximize the signal-to-interference plus noise ratio (SINR). As MF, ZF and MMSE receivers involve large matrix inversion and multiplication operations, they demand a large computational complexity that makes them inefficient for practical implementations. As an alternative solution, Datta et al, \cite{GFDM_IC}, take a time domain successive interference cancellation approach. This solution can completely remove the effect of the self-interference. However, that solution is a computationally exhaustive procedure. In a more recent work from the same research group, Gaspar et al, \cite{GFDM_Rx}, take advantage of the sparsity of the pulse shaping filter in frequency domain to perform the interference cancellation in the frequency domain and hence further reduce the computational complexity of the receiver. Even though the solutions that are based on the results of \cite{GFDM_IC} and \cite{GFDM_Rx} successive interference cancellation can remove the self-interference, they can incur error propagation problems. Recently, Matth\'{e} et al, \cite{GFDM_Gabor}, have proposed a fast algorithm to calculate the ZF and MMSE receiver filters. Their approach is based on the Gabor transform structure of GFDM. Although matrix inversion is circumvented multiplication of the ZF and MMSE matrices to the received signal is a bottle-neck in this approach as the matrix to vector multiplication is a computationally expensive operation.  

In this paper, we design a low complexity transceiver structure for GFDM and therefore improve on the existing approaches. The special structure of the modulation matrix is utilized to reduce the complexity of the transmitter. Compared with the existing GFDM transmitter \cite{GFDM_5G}, so far known to have the lowest complexity, our proposed transmitter structure is more computationally efficient. Based on the lessons that we learned from ICI cancellation in uplink OFDMA systems with interleaved subcarrier allocation, \cite{AF2013}, we are able to substantially reduce the complexity of the ZF and MMSE receivers compared with the low complexity receiver structure that is proposed in \cite{GFDM_Rx}. We propose a unified structure for the MF, ZF and MMSE receivers. This unified receiver structure is beneficial as only the filter coefficients need to be changed for implementation of different receivers. These coefficients can be saved on memory and be used if needed in different scenarios. For instance, ZF receiver can be used instead of MMSE one at high signal-to-noise ratios (SNRs). As our techniques are direct and no approximation is involved, our proposed receivers do not incur any performance loss compared with the optimal MF, ZF and MMSE receivers. Another advantage of our receiver structure with respect to interference cancellation receivers is that it is not iterative and hence the computations can run in parallel which can in turn reduce the overall processing delay of the system. As our proposed transceiver structure is based on sparsification of the matrices that are involved, it also provides savings in the memory requirements of the system.

The rest of the paper is organized as follows. Section~\ref{sec:SystemModel} presents the GFDM system model. Sections~\ref{sec:ProposedTx} and~\ref{sec:ProposedRx} include the design and implementation of our proposed GFDM transmitter and receiver structures, respectively. The computational complexity of our transceiver pair is analyzed in Section~\ref{sec:Complexity}. Finally, the conclusions are drawn in Section~\ref{sec:Conclusion}.

\textit{Notations:} Matrices, vectors and scalar quantities are denoted by boldface uppercase, boldface lowercase and normal letters, respectively. $[\bA]_{m,n}$ and $[\ba]_{n}$ represent the element in the $m^{\rm{th}}$ row and $n^{\rm{th}}$ column of $\bA$ and the $n^{\rm th}$ element of $\ba$, respectively and $\bA^{-1}$ signifies the inverse of $\bA$. $\I_M$ and ${\bf{0}}_M$ are the identity and zero matrices of the size $M\times M$, respectively. $\bD={\rm diag}(\ba)$ is a diagonal matrix whose diagonal elements are formed by the elements of the vector $\ba$ and $\bC={\rm circ}(\ba)$ is a circulant matrix whose first column is $\ba$. The round-down operator $\lfloor\cdot\rfloor$, rounds the value inside to the nearest integer towards minus infinity. The superscripts $(\cdot)^{\rm T}$, $(\cdot)^{\rm H}$ and $(\cdot)^\ast$ indicate transpose, conjugate transpose and conjugate operations, respectively. Finally, $\delta(\cdot)$, $\textcircled{\scriptsize M}$ and ${\rm mod}~N$ represent the Dirac delta function, $M$-point circular convolution and modulo-N operations, respectively.

\section{System Model for GFDM }\label{sec:SystemModel}
\begin{figure*}[ht]
\psfrag{P}[][]{{\scriptsize Circ. Conv.}}
\psfrag{N}[][]{{\scriptsize $~N$}}
\psfrag{C}[][]{{\scriptsize Channel}}
\psfrag{E}[][]{{\scriptsize estimation}}
\psfrag{FDE}[][]{{\scriptsize FDE}}
\psfrag{SFB}[][]{{\scriptsize Synthesis filter bank}}
\psfrag{AFB}[][]{{\scriptsize Analysis filter bank}}
\psfrag{A}[][]{{\scriptsize Addition}}
\psfrag{R}[][]{{\scriptsize removal}}
\psfrag{CP}[][]{{\scriptsize CP}}
\psfrag{g}[][]{$\scriptstyle g_n$}
\psfrag{g1}[][]{$\scriptstyle \breve{g}_n$}
\psfrag{S1}[][]{$\scriptstyle \mathbf{d}_0$}
\psfrag{S2}[][]{$\scriptstyle \mathbf{d}_1$}
\psfrag{SN}[][]{$\scriptstyle \mathbf{d}_{N-1}$}
\psfrag{Sh1}[][]{$\scriptstyle \hat{\mathbf{d}}_0$}
\psfrag{Sh2}[][]{$\scriptstyle \hat{\mathbf{d}}_1$}
\psfrag{ShN}[][]{$\scriptstyle \hat{\mathbf{d}}_{N-1}$}
\psfrag{e1}[][]{{$\scriptstyle e^{\frac{j2\pi n}{N}}$}}
\psfrag{en}[][]{{$\scriptstyle e^{\frac{j2\pi n}{N}(N-1)}$}}
\psfrag{me1}[][]{{$\scriptstyle e^{-\frac{j2\pi n}{N}}$}}
\psfrag{men}[][]{{$\scriptstyle e^{-\frac{j2\pi n}{N}(N-1)}$}}
\centering
\includegraphics[scale=0.25]{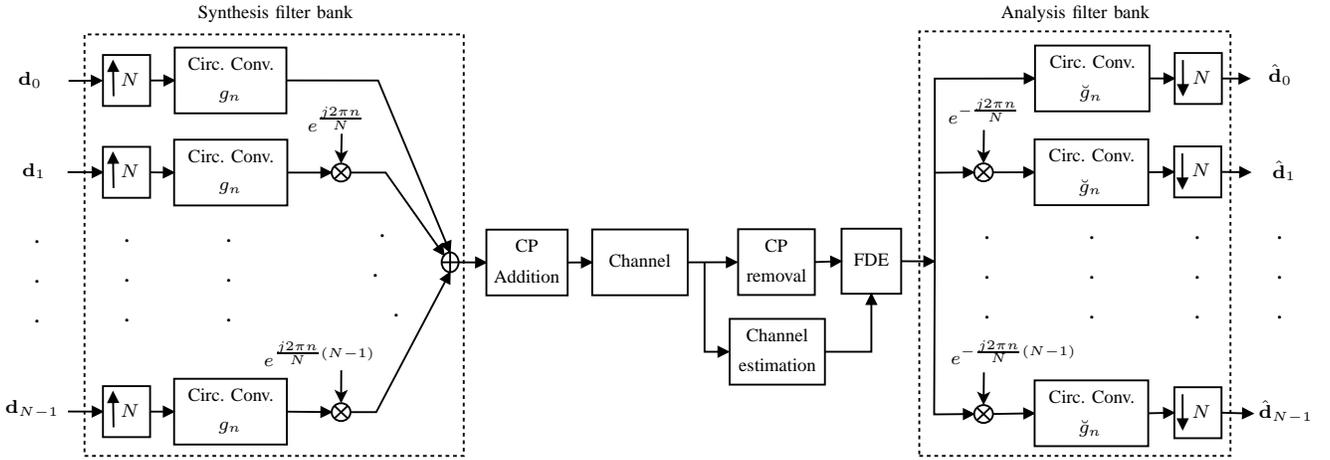}
\caption{Baseband block diagram of a GFDM transceiver system.}
\vspace{-3 mm}
\label{fig:GFDM_Transceiver}
\end{figure*}

We consider a GFDM system with the total number of $N$ subcarriers that includes $M$ symbols in each block. In a GFDM block, $M$ symbols overlap in time. Therefore, we call $M$, \textit{overlapping factor} of the GFDM system. The $MN\times 1$ vector $\dd = [\dd_0^{\rm T},\ldots,\dd_{N-1}^{\rm T}]^{\rm T}$ contains the complex data symbols of the GFDM block where the $M\times 1$ data vector $\dd_i=[d_i(0),\ldots,d_i(M-1)]^{\rm T}$ contains the data symbols to be transmitted on the $i^{\rm th}$ subcarrier. To put it differently, $d_i(m)$ is the data symbol to be transmitted at the $m^{\rm th}$ time slot on the $i^{\rm th}$ subcarrier. The data symbols are taken from a zero mean independent and identically distributed (i.i.d) process with the variance of unity. In GFDM modulation, the data symbols to be transmitted on the $i^{\rm th}$ subcarrier are first up-sampled by the factor of $N$ to form an impulse train
\be\label{eqn:si}
s_i(n)=\sum^{M-1}_{k=0}{d_i(k)\delta(n-kN)},~~n=0,\ldots,NM-1.
\ee
Then, $\bs_i = [s_i(0),\ldots,s_i(MN-1)]^{\rm T}$ is circularly convolved with the prototype filter and up-converted to its corresponding subcarrier frequency. After performing the same procedure for all the subcarriers, the resulting signals are summed up to form the GFDM signal $x(n)$, \cite{GFDM_ber}.
\be\label{eqn:xn}
x(n)=\sum^{N-1}_{i=0}\sum^{M-1}_{m=0}{d_i(m)g_{\{(n-mN)~{\rm mod}~MN\}}e^{j\frac{2\pi in}{N}}},
\ee 
where $g_\ell$ is the $\ell^{\rm th}$ coefficient of the prototype filter.  

Putting together all the transmitter output samples in an $MN\times 1$ vector $\x=[x(0),\ldots,x(MN-1)]^{\rm T}$, the GFDM signal can be represented as multiplication of a modulation matrix ${\bA}$ of size $MN\times MN$ to the data vector ${\dd}$, \cite{GFDM_ber}. 
\be\label{eqn:x}
\bf{x} = \bf{Ad}.
\ee

Modulation matrix ${\bf{A}}$ encompasses all signal processing steps involved in modulation. Let ${\g}=[g_0,\ldots,g_{MN-1}]^{\rm T}$ hold all the coefficients of the pulse shaping/prototype filter with the length $MN$, the elements of ${\bA}$ can be represented as,
\be\label{eqn:A_nm}
[\bA]_{nm} = g_{\{(n-mN)~{\rm mod}~MN\}}e^{j\frac{2\pi n}{N}\left\lfloor{\frac{m}{M}}\right\rfloor}.
\ee
Based on the equations (\ref{eqn:xn}) to (\ref{eqn:A_nm}), the matrix $\bA$ can be written as
\be\label{eqn:Amat}
\bA = 
\begin{bmatrix}
\BG&\BE_1\BG&\hdots&\BE_{N-1}\BG
\end{bmatrix},
\ee
where $\BG$ is an $MN\times M$ matrix whose first column contains the samples of the prototype filter $\bg$ and its consecutive columns are the copies of the previous column circularly shifted by $N$ samples. $\BE_i = {\rm diag}\{[\e_i^{\rm T},\ldots,\e_i^{\rm T}]^{\rm T}\}$ is an $MN\times MN$ diagonal matrix whose diagonal elements are comprised of $M$ concatenated copies of the vector $\e_i=[1,e^{j\frac{2\pi i}{N}},\ldots,e^{j\frac{2\pi i}{N}(N-1)}]^{\rm T}$.

GFDM systems use frequency domain equalization (FDE) to tackle the wireless channel impairments and reduce the channel equalization complexity. In those systems, a CP which is longer than the channel delay spread is added to the beginning of the GFDM block to accommodate the channel transient period. If $N_{\rm CP}$ is the CP length, the last $N_{\rm CP}$ elements of the vector $\x$ are appended to its beginning in order to form the transmitted signal vector $\bar{\x}$ whose length is $MN+N_{\rm CP}$. Let $\bh=[h_0,\ldots,h_{N_{\rm ch}-1}]^{\rm T}$ be the channel impulse response. Thus, the CP length $N_{\rm CP}$ needs to be longer than the channel length $N_{\rm ch}$. The received signal which has gone through the channel, after CP removal can be shown as
\be\label{eqn:received}
\br= \bH\x +{\boldsymbol{\nu}},
\ee 
where ${\boldsymbol{\nu}}$ is the complex additive white Gaussian noise (AWGN) vector, i.e, $\boldsymbol\nu~{\sim}~{\mathcal {CN}}(0,{{\sigma_\nu}^{2}}{\I_{MN}})$, ${\sigma_\nu}^{2}$ is the noise variance, $\bH={\rm circ}\{\tilde{\bh}\}$ and $\tilde{\bh}$ is the zero padded version of $\bh$ to have the same length as $\x$. Due to the fact that $\bH$ is a circulant matrix, an FDE procedure can be performed to compensate for the multipath channel impairments. With the assumption of having perfect synchronization and channel estimates, the equalized signal can be obtained as
\be\label{eqn:eqRx}
\y = \bF^{\rm H}_{MN}\BH^{-1}\bF_{MN}\br,
\ee
where $\bF_{MN}$ is $MN$-point normalized discrete Fourier transform (DFT) matrix and $\BH^{-1}$ is a diagonal matrix whose diagonal elements are reciprocals of the elements of the vector obtained from taking $MN$-point DFT of the zero padded version of $\bh$, viz., $\tilde{\bh}$. The vector $\y=[y_0,\ldots,y_{MN-1}]^{\rm T}$ is the output of the FDE block. 

In order to suppress or remove the ICI due to non-orthogonality of the subcarriers and estimate the transmitted data vector $\bd$ from the equalized signal vector,  three linear GFDM receivers; namely, MF, ZF and MMSE detectors are considered in this paper.

As it was discussed in, \cite{GFDM_ber}, the transmitted symbols can be recovered through match filtering
\be\label{eqn:MF}
\hat{\bd}_{\rm MF} = \bA^{\rm H}\y.
\ee 
However, MF receiver cannot completely remove the ICI. Hence, ZF solution can be utilized to completely eliminate the ICI that is caused by non-orthogonality of the subcarriers. The ZF estimate of the transmitted data vector can be found as
\be\label{eqn:ZF}
\hat{\bd}_{\rm ZF} = (\bA^{\rm H}\bA)^{-1}\bA^{\rm H}\y.
\ee 
Since $(\bA^{\rm H}\bA)^{-1}\bA^{\rm H}$ can have large values, its multiplication to $\y$ can result in noise enhancement. This noise amplification problem can be taken care of by utilizing the MMSE receiver
\be\label{eqn:MMSE}
\hat{\bd}_{\rm MMSE} = (\bA^{\rm H}\bA+{\sigma_\nu}^{2}\I_{MN})^{-1}\bA^{\rm H}\y.
\ee 

Fig.~\ref{fig:GFDM_Transceiver}, depicts the baseband block diagram of a GFDM transceiver when we have perfect synchronization in time and frequency between the transmitter and receiver. Fig.~\ref{fig:GFDM_Transceiver} summarizes the modulation and demodulation process that is discussed above. It is worth mentioning that $g_n$'s for $n=0,\ldots,MN-1$ are the prototype filter coefficients and $\breve{g}_n$'s are the receiver filter coefficients which can be taken from the coefficients of MF, ZF or MMSE receiver filter. As it was mentioned in Section~\ref{sec:Intro}, GFDM is a type of filter bank multicarrier system with circular pulse shaping. Therefore, GFDM transmitter and receiver can be thought of as a pair of synthesis and analysis filter banks, respectively.    
\begin{figure*}[ht]
\psfrag{C}[][]{{\scriptsize Circular Convolution}}
\psfrag{G0}[][]{$\scriptstyle\bar{\g}_0$}
\psfrag{G1}[][]{$\scriptstyle\bar{\g}_1$}
\psfrag{G3}[][]{$\scriptstyle\bar{\g}_{N-1}$}
\psfrag{yt}[][]{$d_n$}
\psfrag{yt1}[][]{$\bar{d}_n$}
\psfrag{z}[][]{$z^{-1}$}
\psfrag{P}[][]{{\scriptsize $N$-point}}
\psfrag{D}[][]{{\scriptsize DFT}}
\psfrag{O}[][]{$x_n$}
\psfrag{M}[][]{{$~M$}}
\psfrag{TXT}[][]{\tiny{This commutator}}
\psfrag{TXT1}[][]{\tiny{turns by one position}}
\psfrag{TXT2}[][]{\tiny{after every $M$ samples}}
\psfrag{PS}[][]{\scriptsize{P/S}}
		\centering 
		\subfigure[]{
    \includegraphics[scale=0.24]{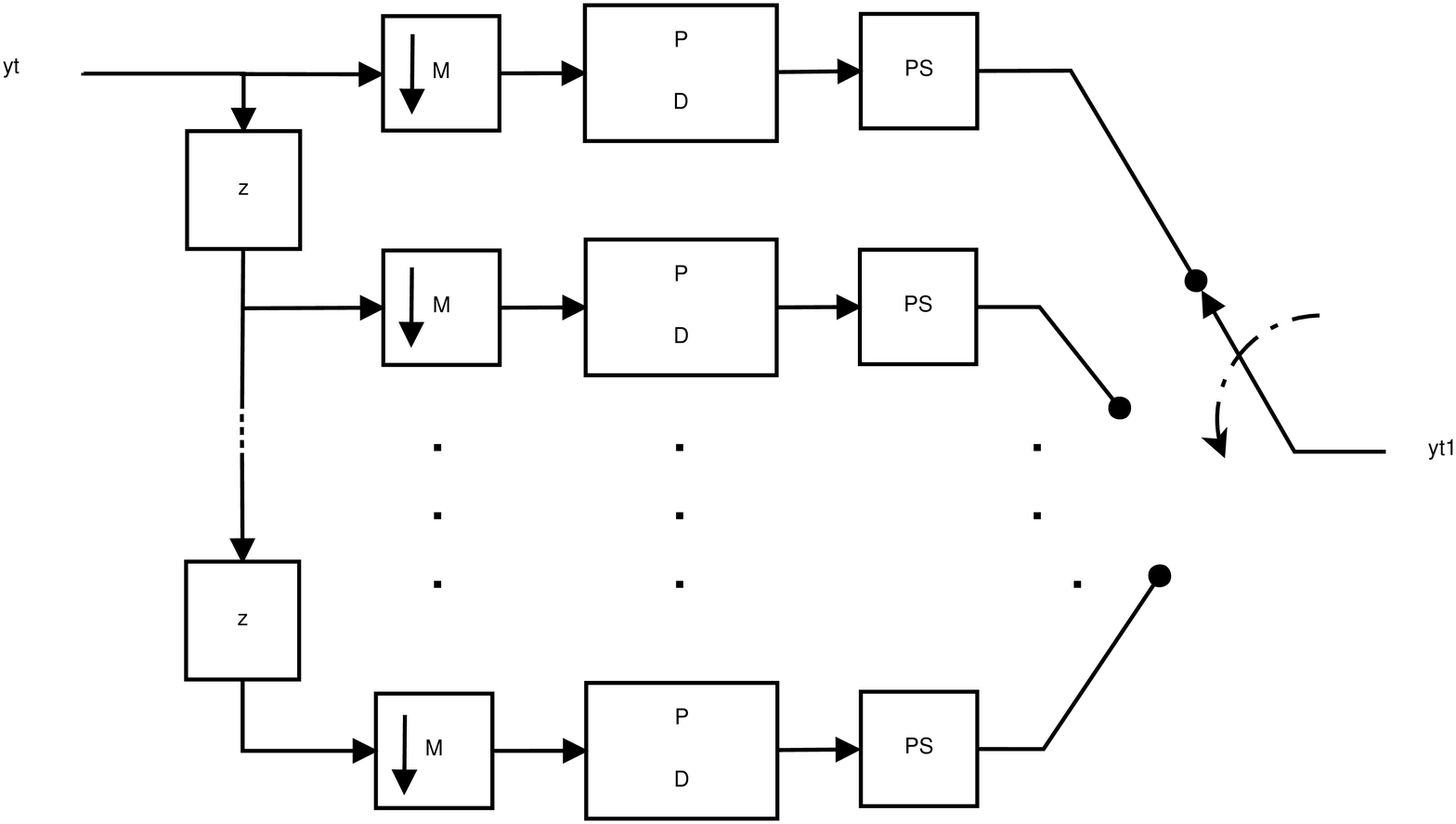}
    \label{subfig:dnhat}
    }
	\subfigure[]{	
   \includegraphics[scale=0.24]{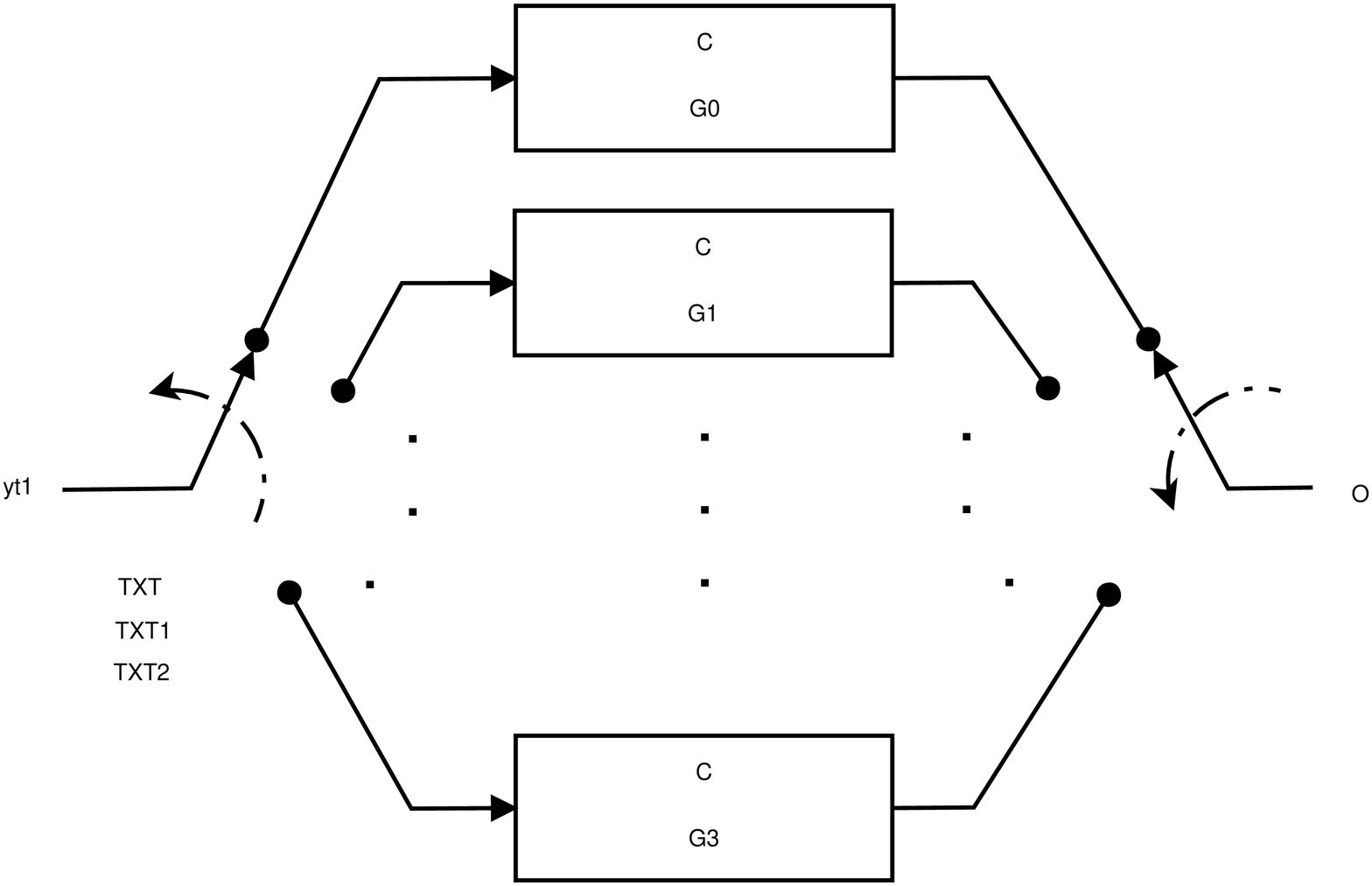}
    \label{subfig:x}
	}
\caption{Concatenation of \subref{subfig:yt} and \subref{subfig:dhat} show the implementation of the proposed GFDM transmitter.}
\vspace{-2 mm}
\label{fig:Tx}
\end{figure*} 

From equations (\ref{eqn:x}) and (\ref{eqn:MF}) to (\ref{eqn:MMSE}), one realizes that direct matrix multiplications and inversions that are involved, demand a very large computational complexity as all the matrices are of the size $MN\times MN$, with $N$ being usually large, and such complexity may not be affordable for practical systems. Therefore, in the remainder of this paper, low complexity techniques will be proposed that can substantially reduce the computational cost of the synthesis and analysis filter banks that are shown in Fig.~\ref{fig:GFDM_Transceiver}, while maintaining the optimal performance.

\section{Proposed GFDM Transmitter}\label{sec:ProposedTx} 
This section presents our proposed low complexity GFDM transmitter design and implementation. In the following subsections, we will show how the synthesis filter bank of Fig.~\ref{fig:GFDM_Transceiver} can be simplified to have a very low computational load.
\subsection{GFDM transmitter design}\label{subsec:TxDesign}
Starting from (\ref{eqn:x}), one can realize that direct multiplication of the matrix $\bA$ to the data vector $\bd$ is a complex operation which demands $(MN)^2$ complex multiplications. Therefore, complexity will be an issue for practical systems as the number of subcarriers and/or the parameter $M$ increases. Accordingly, a low complexity implementation technique for GFDM transmitter has to be sought. To this end, equation (\ref{eqn:x}) can be written as
\be\label{eqn:x1}
\x = \bA\bd = \bA\BDFT_b^{\rm H}\BDFT_b\bd,
\ee  
where $\BDFT_b$ is the $MN\times MN$ normalized block DFT matrix that includes $M\times M$ submatrices $\bOmega_{ni}=\frac{1}{\sqrt{N}}e^{-j\frac{2\pi ni}{N}}\I_M$ and $n,i=0,\ldots,N-1$. Validity of equation (\ref{eqn:x1}) is based on the fact that $\BDFT_b^{\rm H}\BDFT_b=\I_{MN}$. As it is derived in Appendix~\ref{appendix:FbAh}, the resulting matrix from multiplication of the block DFT matrix $\BDFT_b$ into $\bA^{\rm H}$ is sparse and it is comprised of the prototype filter coefficients scaled by $\sqrt{N}$. From equation (\ref{eqn:x1}), it can be inferred that $\bGamma^{\rm H}=\bA\BDFT_b^{\rm H}$ is also sparse since it is the conjugate transpose of $\BDFT_b\bA^{\rm H}$. Hence, our strategy allows us to make the matrix $\bA$ sparse and real as the prototype filter is usually chosen as a real filter. Due to (\ref{eqn:x1}) and the definition of $\BDFT_b$, $\BDFT_b\bd$ can be implemented by performing $M$ DFT operations of size $N$ on the data samples, i.e., one per GFDM symbol. Let $\bar{\bd}=\BDFT_b\bd=[\bar{\bd}^{\rm T}_0,\ldots,\bar{\bd}^{\rm T}_{N-1}]^{\rm T}$ where the $M\times 1$ vector $\bar{\bd}_i=[\bar{d}_i(0),\ldots,\bar{d}_{i}(M-1)]^{\rm T}$ contains the $i^{\rm th}$ output of each DFT block, then (\ref{eqn:x1}) can be rearranged as
\be\label{eqn:x2}
\x = \bGamma^{\rm H}\bar{\bd}=\sum_{i=0}^{N-1}{\bGamma_i^{\rm H}\bar{\bd}_\kappa},
\ee
where $\kappa = {(N-i)~{\rm mod}~N}$. As discussed in Appendix~\ref{appendix:DClosedForm}, the $M\times MN$ matrices $\bGamma_i$'s have only $M$ non-zero columns and the sets of those column indices are mutually exclusive with respect to each other. As a result, ${\bGamma_i^{\rm H}\bar{\bd}_\kappa}$ will be a sparse vector with only $M$ non-zero elements located on the positions $\kappa,\kappa+N,\ldots,\kappa+(M-1)N$. On the basis of the derivations that are presented in Appendix~\ref{appendix:FbAh}, the non-zero elements of ${\bGamma_i^{\rm H}\bar{\bd}_\kappa}$ can be obtained from $M$-point circular convolution of $\bar{\bd}_\kappa$ with the $\kappa^{\rm th}$ polyphase component of the prototype filter $\g_\kappa$ that is scaled by $\sqrt{N}$. Therefore, defining the non-zero elements of $\bGamma_i^{\rm H}\bar{\bd}_\kappa$ as the vector $\x_\kappa=[x_\kappa,x_{\kappa+N},\ldots,x_{\kappa+(M-1)N}]^{\rm T}$, we get
\be\label{eqn:x3}
\x_\kappa = \bar{\g}_\kappa\textcircled{\scriptsize M}\bar{\bd}_\kappa,
\ee
where $\bar{\g}_\kappa=\sqrt{N}\g_\kappa$.
\subsection{GFDM transmitter implementation}\label{subsec:TxImplementation}
In this subsection, implementation of the designed GFDM transmitter in Section~\ref{subsec:TxDesign} is discussed. From the equations (\ref{eqn:x1}) to (\ref{eqn:x3}), GFDM modulation, based on our design, can be summarized into two steps.
\begin{enumerate}
	\item $M$ number of $N$-point DFT operations, i.e., application of $N$-point DFT to each individual GFDM symbol which includes $N$ subcarriers. This can be efficiently implemented by taking advantage of the fast Fourier transform (FFT) algorithm.
	\item $N$ number of $M$-point circular convolution operations. 
\end{enumerate}

Therefore, the first and second steps of our GFDM transmitter can be implemented by cascading the block diagrams shown in Fig.~\ref{fig:Tx} \subref{subfig:dnhat} and \subref{subfig:x}, respectively. The blocks P/S convert the parallel FFT outputs to serial streams. All the commutators shown in Fig.~\ref{fig:Tx} turn counter clockwise. Both commutators located on the right hand side of the Fig.~\ref{fig:Tx} \subref{subfig:dnhat} and \subref{subfig:x} turn after one sample collection. However, the one located on the left hand side of \subref{subfig:x} turns by one position after sending $M$ samples to each $M$-point circular convolution block. 

\section{Proposed GFDM Receiver}\label{sec:ProposedRx} 
In this section, we derive low complexity ZF and MMSE receivers for GFDM systems. It is worth mentioning that our solutions are direct and hence lower complexity of these receivers comes for free as they do not result in any performance loss, thanks to the special structure of the matrix $\bA^{\rm H}\bA$. The characteristics of $\bA^{\rm H}\bA$ will be discussed in the next subsection and then we will derive our proposed receivers on the basis of those traits.

\subsection{Block-diagonalization of the matrix $\bA^{\rm H}\bA$}\label{subsec:Bdiagonalization}
The key idea behind our proposed GFDM receiver techniques is to take advantage of the particular structure of the matrix $\bA^{\rm H}\bA$ which is present in both ZF and MMSE receiver formulations. Using (\ref{eqn:Amat}), one can calculate $\bA^{\rm H}\bA$ and find out that it has the following structure
\be\label{eqn:AHA}
 \bA^{\rm H}\bA =
 \begin{bmatrix}
  \BG^{\rm H}\BG & \BG^{\rm H}\BE_1\BG & \cdots & \BG^{\rm H}\BE_{N-1}\BG \\
  \BG^{\rm H}\BE_1^{\rm H}\BG & \BG^{\rm H}\BG & \cdots & \BG^{\rm H}\BE_{N-2}\BG \\
  \vdots  & \vdots  & \ddots & \vdots  \\
  \BG^{\rm H}\BE_{N-1}^{\rm H}\BG & \BG^{\rm H}\BE_{N-2}^{\rm H}\BG & \cdots & \BG^{\rm H}\BG
 \end{bmatrix}.
\ee
From the definition of vector $\e_i$, it can be straightforwardly perceived that $\e_{N-i}^{\rm H}=\e_i$ and hence $\BE_{N-i}^H=\BE_i$. Therefore, the columns of $\bA^{\rm H}\bA$ as shown in (\ref{eqn:AHA}) are circularly shifted with respect to each other. Accordingly, $\bA^{\rm H}\bA$ is a block-circulant matrix with blocks of size $M\times M$. Following a similar line of derivations as in \cite{Gerlic1983} and \cite{AF2013}, $\bA^{\rm H}\bA$ can be expanded as follows
\be\label{eqn:AHAexp}
\bA^{\rm H}\bA=\BDFT_b^{\rm H}\BD\BDFT_b,
\ee
where $\BD$ is an $MN\times MN$ block-diagonal matrix, $\BD={\rm diag}\{\BD_0,\ldots,\BD_{N-1}\}$ and $\BD_i$'s are $M\times M$ block matrices. From (\ref{eqn:AHAexp}), $\BD$ can be derived as
\be\label{eqn:D}
\BD=\BDFT_b(\bA^{\rm H}\bA)\BDFT_b^{\rm H}.
\ee
As it is explained in Appendix~\ref{appendix:DClosedForm}, $\BD_i$'s can be derived from polyphase components of the prototype filter.
\be\label{eqn:Dis0}
\BD_i = N{\rm circ}\{{\g}_{\kappa}\textcircled{\scriptsize M}\tilde{\g}_{\kappa}\},
\ee  
where $\kappa = {(N-i)~{\rm mod}~N}$, $\g_i$ is the $i^{\rm th}$ polyphase component of $\g$ and $\tilde{\g}_{i}=[g_i,g_{i+(M-1)N},\ldots,g_{i+N}]^{\rm T}$ is its circularly folded version. As (\ref{eqn:Dis0}) highlights, $\BD_i$'s are all real and circulant matrices.  
\begin{figure*}[ht]
\psfrag{C}[][]{{\scriptsize Circular Convolution}}
\psfrag{G0}[][]{$\bgamma_0$}
\psfrag{G1}[][]{$\bgamma_1$}
\psfrag{G3}[][]{$\bgamma_{N-1}$}
\psfrag{y}[][]{$y_n$}
\psfrag{yt}[][]{\hspace{5 mm}$\bar{y}_n/\tilde{y}_n/\breve{y}_n$}
\psfrag{ytt}[][]{$\bar{y}_n/\tilde{y}_n/\breve{y}_n$\hspace{5 mm}}
\psfrag{N}[][]{{\small$~N$}}
\psfrag{z}[][]{$z^{-1}$}
\psfrag{P}[][]{{\scriptsize $N$-point}}
\psfrag{D}[][]{{\scriptsize IDFT}}
\psfrag{O}[][]{$\hat{d}_n$}
\psfrag{M}[][]{{$~M$}}
\psfrag{TXT}[][]{\tiny{This commutator}}
\psfrag{TXT1}[][]{\tiny{turns by one position}}
\psfrag{TXT2}[][]{\tiny{after collecting $M$ samples}}
\psfrag{PS}[][]{\scriptsize{P/S}}
		\centering 
		
		\subfigure[]{
    \includegraphics[scale=0.24]{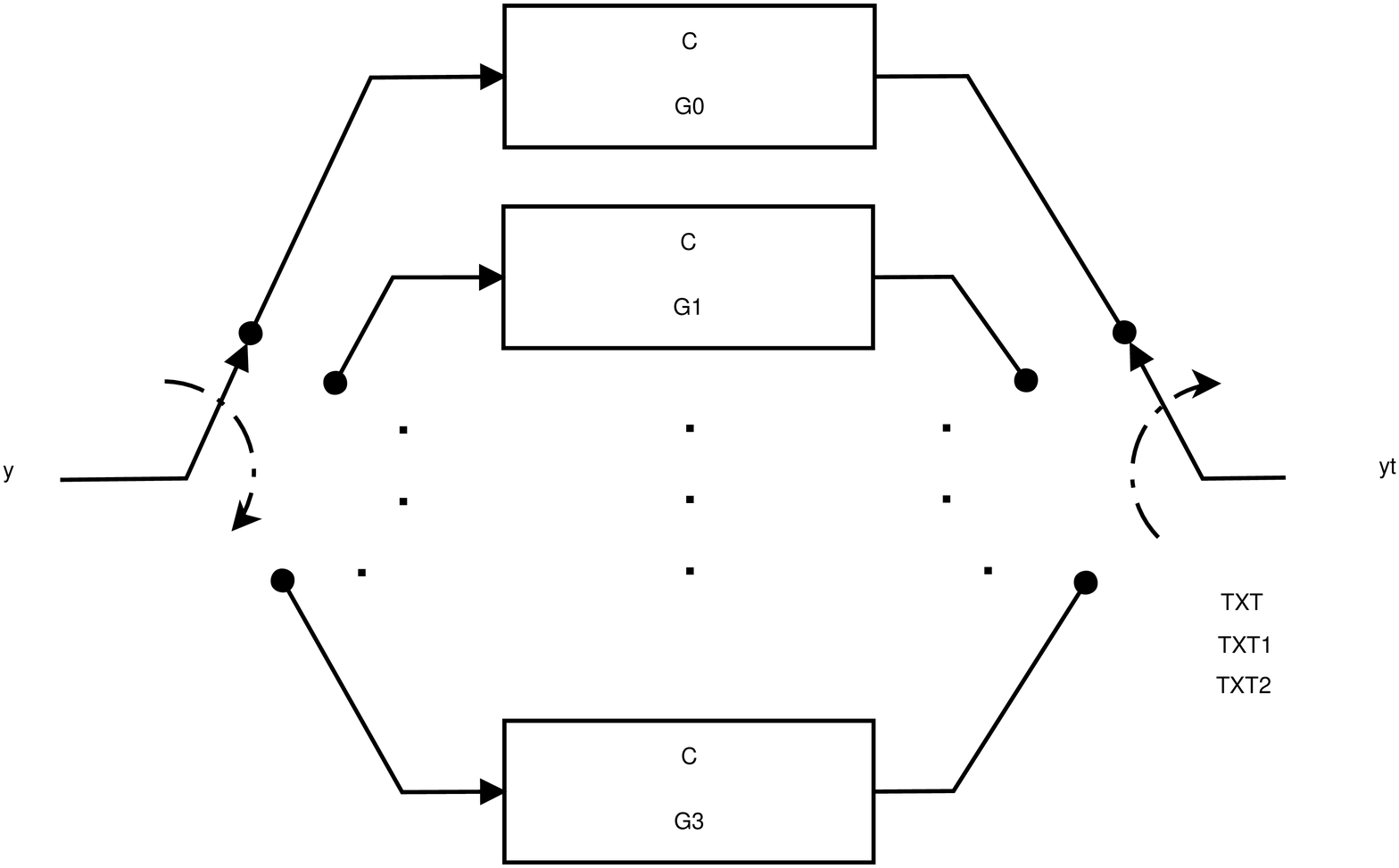}
    \label{subfig:yt}
    }
	\subfigure[]{
	\includegraphics[scale=0.24]{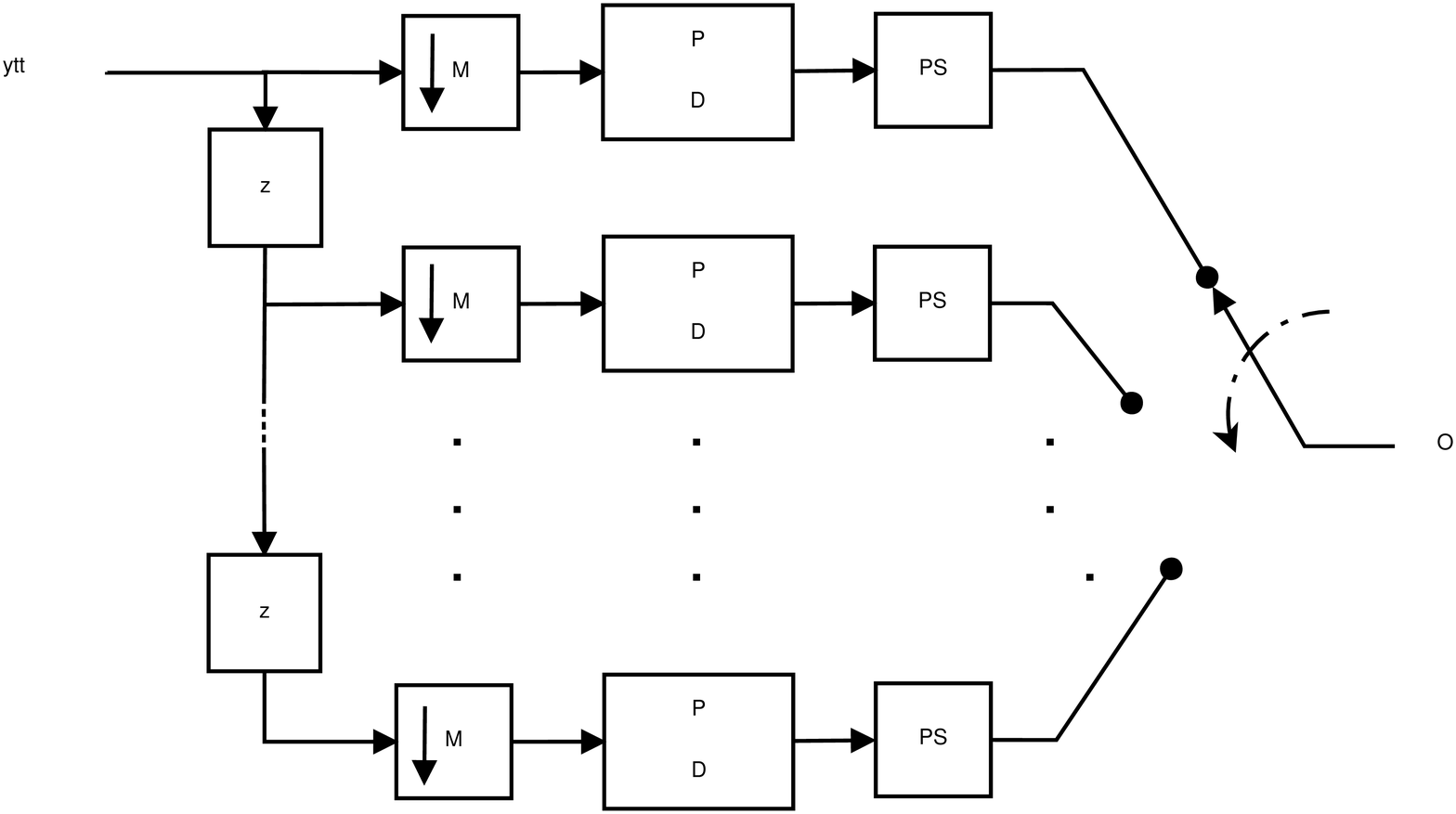}
    \label{subfig:dhat}
	}
\caption{Unified implementation of our proposed MF, ZF and MMSE-based GFDM receivers from cascading the block diagrams \subref{subfig:yt} and \subref{subfig:dhat}.}
\vspace{-2 mm}
\label{fig:BlockDiagram}
\end{figure*}
\subsection{Low complexity MF receiver}\label{subsec:MF_Rx}
Based on equation (\ref{eqn:MF}), direct implementation of MF receiver involves a matrix to vector multiplication which has the computational cost of $(MN)^2$ complex multiplications. This procedure becomes highly complex for large values of $N$ and/or $M$ which is usually the case. As discussed in Appendix~\ref{appendix:FbAh}, multiplication of $\bA^{\rm H}$ by the block DFT matrix results in a sparse matrix. Due to the fact that $\BDFT_b^{\rm H}\BDFT_b=\I_{MN}$, similar to the transmitter (equation(\ref{eqn:x1})), equation (\ref{eqn:MF}) can be written as
\begin{eqnarray}\label{eqn:MF2}
\hat{\bd}_{\rm MF} &=& \BDFT_b^{\rm H}\BDFT_b\bA^{\rm H}\y \nonumber\\
&=&\BDFT_b^{\rm H}\bGamma\y,
\end{eqnarray}
where $\bGamma$ is a sparse matrix with only $NM^2$ non-zero elements that are the scaled version of the prototype filter coefficients. Closed form of $\bGamma  = [\bGamma_0^{\rm T},\ldots,\bGamma_{N-1}^{\rm T}]^{\rm T}$ is derived in Appendix~\ref{appendix:FbAh} and it is shown that the matrix is real valued and comprised of the prototype filter elements. Non-zero columns of the $M\times MN$ block matrices $\bGamma_i$'s are circularly shifted copies of each other. Hence, multiplication of $\bGamma_i$ and $\y$ is equivalent to $M$-point circular convolution of $M$ equidistant elements of $\y$ starting from the $\kappa^{th}$ position and circularly folded version of the $\kappa^{\rm th}$ polyphase component of $\bg$ scaled by $\sqrt{N}$, viz., $\sqrt{N}\tilde{\g}_{\kappa}$. Usually, the prototype filter coefficients are real-valued. Thus, $\bGamma$ is real-valued.  Multiplication of $\BDFT_b^{\rm H}$ to the vector $\bGamma\y$ can be implemented by applying $M$ number of $N$-point IDFT operations. Let $\bar{\y}=\bGamma\y=[\bar{\y}_0^{\rm T},\ldots,\bar{\y}_{N-1}^{\rm T}]^{\rm T}$ and $\y_\kappa=[y_\kappa,y_{\kappa+N},\ldots,y_{\kappa+(M-1)N}]^{\rm T}$. Therefore, we have
\be\label{eqn:MF2}
\bar{\y}_i=\bGamma_i\y={\bv}_\kappa\textcircled{\scriptsize M}\y_\kappa,
\ee
where ${\bv}_\kappa=\sqrt{N}\tilde{\g}_\kappa$. Finally, the MF estimates of $\bd$ can be obtained as
\be\label{eqn:MF3}
\hat{\bd}_{\rm MF}=\BDFT_b^{\rm H}\bar{\y}.
\ee
\subsection{Low complexity ZF receiver}\label{subsec:ZF_Rx}
Inserting (\ref{eqn:AHAexp}) into (\ref{eqn:ZF}), we get
\be\label{eqn:ZFimp}
\hat{\bd}_{\rm ZF} = \BDFT_b^{\rm H}\BD^{-1}\BDFT_b\bA^{\rm H}\y.
\ee
Multiplication of matrix $\bA^{\rm H}$ to the vector $\y$ is the first source of computational burden in ZF receiver which has computational cost of $(MN)^2$. However, this complexity can be reduced by taking advantage of the sparsity of the matrix $\bGamma=\BDFT_b\bA^{\rm H}$ as it was suggested in the previous subsection. Equation (\ref{eqn:MF2}) can be written as $\bar{\y}_i=\bGamma_i\y=\sqrt{N}{\rm circ}\{\tilde{\g}_{\kappa}\}\y_\kappa$. Let $\tilde{\y}=\BD^{-1}\bar{\y}=[\tilde{\y}_0^{\rm T},\ldots,\tilde{\y}_{N-1}^{\rm T}]^{\rm T}$ where 
\be\label{eqn:ytilda}
\tilde{\y}_i=\sqrt{N}\BD_i^{-1}{\rm circ}\{\tilde{\g}_{\kappa}\}\y_\kappa.
\ee
Therefore, from rearranging equation (\ref{eqn:Dis0}) as $\BD_i=N{\rm circ}\{\tilde{\g}_{\kappa}\}{\rm circ}\{{\g}_{\kappa}\}$ and inserting it into (\ref{eqn:ytilda}), we have
\begin{eqnarray}\label{eqn:yitilda}
\tilde{\y}_i&=&\frac{1}{\sqrt{N}}({\rm circ}\{\tilde{\g}_{\kappa}\}{\rm circ}\{{\g}_{\kappa}\})^{-1}{\rm circ}\{\tilde{\g}_{\kappa}\}\y_\kappa \nonumber \\
&=&\frac{1}{\sqrt{N}}({\rm circ}\{\g_{\kappa}\})^{-1}\y_\kappa \nonumber \\
&=&\q_\kappa\textcircled{\scriptsize M}\y_\kappa,
\end{eqnarray}
where $\q_\kappa$ includes the first column of the circulant matrix $({\rm circ}\{\g_{\kappa}\})^{-1}$ scaled by $\frac{1}{\sqrt{N}}$. Due to the fact that the the coefficients of the prototype filter are known, the vectors $\q_\kappa$'s can be calculated offline. Additionally, since the prototype filter coefficients are real, $\q_\kappa$'s are also real. From (\ref{eqn:yitilda}), one may realize that calculation of the vector $\tilde{\y}$ needs $N$ number of $M$-point circular convolutions. After acquiring $\tilde{\y}$, the ZF estimates of the transmitted symbols can be obtained as
\be\label{eqn:ZF1}
\hat{\bd}_{\rm ZF} = \BDFT_b^{\rm H}\tilde{\y}.
\ee
As can be inferred from (\ref{eqn:ZF1}), finding $\hat{\bd}_{\rm ZF}$ from $\tilde{\y}$ requires $M$ number of $N$-point inverse DFT (IDFT) operations. 

\subsection{Low complexity MMSE receiver}\label{subsec:MMSE_Rx}
Using (\ref{eqn:AHAexp}) in (\ref{eqn:MMSE}) we get
\begin{eqnarray}\label{eqn:MMSE0}
\hat{\bd}_{\rm MMSE} &=& (\BDFT_b^{\rm H}\BD\BDFT_b+{\sigma_\nu}^{2}\I_{MN})^{-1}\bA^{\rm H}\y \nonumber \\
&=& \BDFT_b^{\rm H}\bar{\BD}^{-1}\BDFT_b\bA^{\rm H}\y,
\end{eqnarray}
where $\bar{\BD}=\BD+{\sigma_\nu}^{2}\I_{MN}={\rm diag}\{\bar{\BD}_0,\ldots,\bar{\BD}_{N-1}\}$ and $\bar{\BD}_i = {\BD}_i+{\sigma_\nu}^{2}\I_{M}$. Recalling circulant property of $\BD_i$ from (\ref{eqn:Dis0}), it can be understood that $\bar{\BD}_i$ is also circulant and can be expanded as $\bar{\BD}_i=\bF^{\rm H}_M(\bPhi_\kappa^\ast\bPhi_\kappa+{\sigma_\nu}^{2}\I_{M})\bF_M$ where $\bPhi_\kappa=MN{\rm diag}\{\bF_M\g_\kappa\}$\footnote{Since, $\tilde{\g}_\kappa$ is a real vector and circularly folded version of ${\g}_\kappa$, $\bPhi_\kappa^\ast=MN{\rm diag}\{\bF_M\tilde{\g}_\kappa\}$.}. Let $\breve{\y}=[\breve{\y}_0^{\rm T},\ldots,\breve{\y}_{N-1}^{\rm T}]^{\rm T}=\bar{\BD}^{-1}\BDFT_b\bA^{\rm H}\y$, we can write
\begin{eqnarray}\label{eqn:ybreve}
\breve{\y}_i &=& \bF^{\rm H}_M(\bPhi_\kappa^\ast\bPhi_\kappa+{\sigma_\nu}^{2}\I_{M})^{-1}\bPhi_\kappa^\ast\bF_M\y_\kappa \nonumber \\
&=& \p_\kappa\textcircled{\scriptsize M}\y_\kappa,
\end{eqnarray}
where $\p_\kappa$ includes the first column of the circulant matrix $\bF^{\rm H}_M\{(\bPhi_\kappa^\ast\bPhi_\kappa+{\sigma_\nu}^{2}\I_{M})^{-1}\bPhi_\kappa^\ast\}\bF_M$. Since, in MMSE receiver, the matrix $\bar{\BD}^{-1}$ depends on ${\sigma_\nu}^{2}$ and the receiver cannot be simplified as in (\ref{eqn:MF2}) or (\ref{eqn:yitilda}), circular convolution of (\ref{eqn:ybreve}) needs to be calculated in the frequency domain, known as fast convolution, in order to have the lowest complexity. After obtaining $\breve{\y}$, the MMSE estimates of the transmitted symbols can be found as
\be\label{eqn:MMSE1}
\hat{\bd}_{\rm MMSE} = \BDFT_b^{\rm H}\breve{\y}.
\ee
\subsection{Receiver implementation}\label{subsec:RxImp} 
In this subsection, we present a unified implementation of the MF, ZF and MMSE receivers that we proposed in Sections~\ref{subsec:MF_Rx}, \ref{subsec:ZF_Rx} and \ref{subsec:MMSE_Rx}. As Fig.~\ref{fig:BlockDiagram} depicts, the proposed GFDM receivers can be implemented by cascading Fig.~\ref{fig:BlockDiagram}~\subref{subfig:yt} and \subref{subfig:dhat}. It is worth mentioning that the commutator on the right hand side of Fig.~\ref{fig:BlockDiagram}~\subref{subfig:yt} will turn by one position after collecting $M$ samples from the $i^{\rm th}$ branch, i.e., $M\times 1$ vector $\bar{\y}_i/\tilde{\y}_i/\breve{\y}_i$, in the clockwise direction. In the MF and ZF receivers, the vectors $\bgamma_i$ are replaced by ${\bv}_i$'s and $\q_i$'s, respectively, and in MMSE receiver, they will be replaced by $\p_i$'s. Due to the fact that in the MF and ZF receivers, the vectors $\bv_i$ and $\q_i$ are fixed and only depend on the prototype filter coefficients, they can be calculated offline and hence there is no need for their real-time calculation. However, in MMSE receivers, the vectors $\p_i$ depend on the signal to noise ratio and hence they should be calculated in real-time. As mentioned earlier in Section~\ref{subsec:MMSE_Rx}, circular convolutions in our MMSE receiver need to be performed by taking advantage of fast convolution to keep the complexity low.

\section{Computational Complexity}\label{sec:Complexity}
In this section, the computational complexity of our proposed GFDM transmitter and receiver structures are discussed and compared to the existing ones that are known to have the lowest complexity, \cite{GFDM_5G,GFDM_Rx}. In both cases, total number of $N$ subcarriers and overlapping factor of $M$ are considered.

\subsection{Transmitter complexity}\label{subsec:TxComplexity}
Table~\ref{tab:1} presents the computational complexity of different GFDM transmitter implementations based on the number of complex multiplications (CMs). 

As discussed in Section~\ref{subsec:TxImplementation}, our proposed GFDM transmitter involves two steps. The first step includes $M$ number of $N$-point FFT operations that requires $\frac{MN}{2}\log_2N$ CMs. The second step needs $N$ number of $M$-point circular convolutions. Recalling equation (\ref{eqn:x3}), since $\g_\kappa$'s are real-valued vectors, one may realize that each $M$-point circular convolution demands $\frac{M^2}{2}$ number of CMs. If $M$ is a power of two, the complexity can be further reduced by performing the circular convolutions in frequency domain. This is due to the fact that circular convolution in time is multiplication in the frequency domain. Thus, to perform each circular convolution, a pair of $M$-point FFT and IFFT blocks together with $M$ complex multiplications to the filter coefficients in frequency domain are required.

\begin{figure}[t]
		\centering
    \includegraphics[scale=0.6]{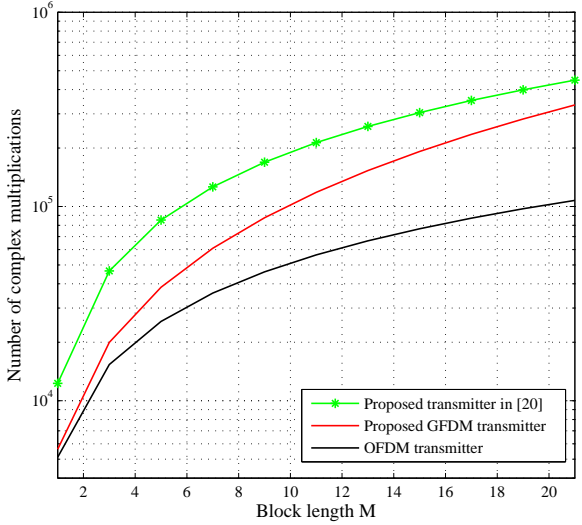}
    \vspace{-7 mm}\caption{Computational complexity comparison of different GFDM transmitter techniques and the OFDM transmitter technique for $N = 1024$.}
\label{fig:TxComplexity}
\vspace{-3 mm}
\end{figure}

The complexity relationships that are presented in Table~\ref{tab:1} are calculated and plotted in Fig.~\ref{fig:TxComplexity} for $N=1024$ subcarriers with respect to different values of overlapping factor $M$. As the authors of \cite{GFDM_5G} suggest, $L=2$ is chosen for calculating their GFDM transmitter complexity. Due to the fact that direct multiplication of $\bA$ to the data vector $\bd$ demands a large number of CMs and is impractical, we do not present it in Fig.~\ref{fig:TxComplexity}. To give a quantitative indication of the complexity reduction that our proposed transmitter provides compared with the direct computation of the equation (\ref{eqn:x}), in the same system setting as used for our other comparisons, i.e., $N=1024$ and $M\in[1,21]$, complexity reduction of around three orders of magnitude can be achieved. According to Fig.~\ref{fig:TxComplexity}, for the small values of $M$ our proposed transmitter structure has a complexity very close to that of OFDM. However, as $M$ increases the complexity of our transmitter increases with a higher pace than OFDM. This is due to the overhead of $\frac{NM^2}{2}$ number of CMs compared with OFDM. Compared with the transmitter structure that we are proposing in this paper, for small values of $M$ up to $11$, the transmitter proposed in \cite{GFDM_5G} demands about two times higher number of CMs. As $M$ increases, complexity of our technique gets close to that of the one proposed in \cite{GFDM_5G}. GFDM transmitter of \cite{GFDM_5G} is about $3$ to $4$ times more complex than OFDM. 

\renewcommand{\arraystretch}{1.5}
\begin{table}[t]
  \centering
    \caption{Computational Complexity of Different GFDM Transmitter Implementations}
    \label{tab:1}
    \resizebox{0.47\textwidth}{!}
{\small \begin{tabular}{|c|c|}
\hline\hline
Technique & Number of Complex Multiplications \\ \hline\hline
Direct matrix multiplication & $(MN)^2$ \\ \hline
Proposed transmitter in~\cite{GFDM_5G} & $MN(\log_2N+2\log_2M+L)$ \\ \hline
Our proposed transmitter & $\frac{MN}{2}(M+\log_2N)$ \\ \hline\hline
    \end{tabular}}
		\vspace{-3 mm}
\end{table}

\subsection{Receiver complexity}\label{subsec:RxComplexity}
Table~\ref{tab:2} summarizes the computational complexity of different GFDM receivers in terms of the number of complex multiplications. The parameter $I$ is the number of iterations in the algorithm with interference cancellation. 

From Fig.~\ref{fig:BlockDiagram}, it can be understood that our proposed receivers involve $N$ and $M$ numbers of $M$-point circular convolutions and $N$-point IDFT operations, respectively. IDFT operations can be efficiently implemented using $N$-point IFFT algorithm which requires $\frac{N}{2}\log_2N$ CMs. As mentioned earlier, in the proposed MF and ZF receivers, the vectors $\bgamma_i$ have fixed values and hence can be calculated and stored offline. Furthermore, $\bgamma_i$'s are real-valued vectors. Thus, the number of complex multiplications needed for $N$ number of $M$-point circular convolutions is $\frac{NM^2}{2}$.

In contrast to the MF and ZF receivers, in the MMSE receiver, the vectors $\bgamma_i$'s are not fixed and depend on the signal-to-noise ratio (SNR). Hence, they need to be calculated in real-time. To this end, as highlighted in Section~\ref{subsec:MMSE_Rx}, those operations can be performed by using $M$-point DFT and IDFT operations. Due to the fact that $(\bPhi_\kappa^\ast\bPhi_\kappa+{\sigma_\nu}^{2}\I_{M})$ is a real-valued diagonal matrix, its inversion and multiplication to $\bPhi_\kappa^\ast$ only needs $\frac{M}{2}$ CMs. The resulting diagonal matrix $(\bPhi_\kappa^\ast\bPhi_\kappa+{\sigma_\nu}^{2}\I_{M})^{-1}\bPhi_\kappa^\ast$ is multiplied into an $M\times 1$ vector which needs $M$ CMs. Since, $M$ is not necessarily a power of $2$, complexity of $M$-point DFT and IDFT operations in the implementation of the circular convolutions is considered as $M^2$. Obviously, if $M$ is a power of $2$, a further complexity reduction by taking advantage of FFT and IFFT algorithms is possible. Therefore, the complexity of our proposed MMSE receiver only differs from the MF and ZF ones in the implementation of the circular convolution operations.

Table~\ref{tab:2} also presents the complexity of the direct MF, ZF and MMSE detection techniques, i.e., direct matrix multiplications and solutions to the equations (\ref{eqn:ZF}) and (\ref{eqn:MMSE}), respectively. Those solutions involve direct inversion of an $MN\times MN$ matrix which has the complexity of $\mathcal{O}(M^3N^3)$ and two vector by matrix multiplications with the computational burden of $2(MN)^2$ CMs.

\begin{figure}[t]
		\vspace{-3 mm}
		\centering
		\includegraphics[scale=0.6]{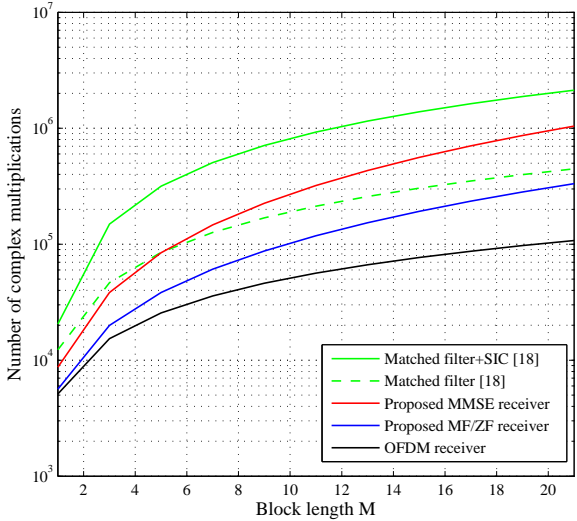}
    \vspace{-7 mm}\caption{Computational complexity comparison of different GFDM receiver techniques with respect to each other and that of OFDM receiver when $N = 1024$ and $I=8$ for \cite{GFDM_Rx}.}
\label{fig:RxComplexity}
\vspace{-4 mm}
\end{figure}

The complexity formulas that are presented in Table~\ref{tab:2} are evaluated and plotted in Fig.~\ref{fig:RxComplexity} for different values of overlapping factor $M\in[1,21]$, $N=1024$ and $I=8$ for the receiver that is proposed in \cite{GFDM_Rx}. Based on the results of \cite{GFDM_Rx}, $I=8$ and $L=2$ are considered. Due to the fact that the complexity of MF, ZF and MMSE receivers with direct matrix inversion and multiplications is prohibitively high compared with other techniques (the difference is in the level of orders of magnitude), they are not presented in Fig.~\ref{fig:RxComplexity}. However, to quantify the amount of complexity reduction that our proposed techniques provide, in the case of $N=1024$ and $M=7$, our proposed MF/ZF receiver is three orders of magnitude and the proposed MMSE receiver is six orders of magnitudes simpler than the direct ones, respectively, in terms of the required number of CMs. As Fig.~\ref{fig:RxComplexity} depicts, our proposed ZF receiver is around an order of magnitude simpler than the proposed receiver with SIC in \cite{GFDM_Rx}. In addition, our proposed MMSE receiver has $2$ to $3$ times lower complexity than the one in \cite{GFDM_Rx}. Apart from lower computational cost compared with the existing receiver structures, our techniques maintain the optimal ZF and MMSE performance as they are direct. Finally, the ZF and MMSE receivers that we are proposing are closer in complexity to OFDM as compared to the receiver in \cite{GFDM_Rx} which is over an order of magnitude more complex than OFDM. 
\renewcommand{\arraystretch}{1.5}
\begin{table}[t]
  \centering
    \caption{Computational Complexity of Different GFDM Receiver Techniques}
    \label{tab:2}
    \resizebox{0.47\textwidth}{!}
{\small \begin{tabular}{|c|c|}
\hline\hline
Technique & Number of Complex Multiplications \\ \hline\hline
Direct ZF & $2(MN)^2$ \\ \hline
Direct MMSE & $\frac{1}{3}(MN)^3+2(MN)^2$ \\ \hline
Matched filter + SIC,~\cite{GFDM_Rx} & $MN(\log_2MN+\log_2M+L+I(2\log_2M+1))$ \\ \hline
Proposed MF/ZF & $\frac{MN}{2}(M+\log_2N)$ \\ \hline
Proposed MMSE & $\frac{MN}{2}(4M+\log_2N+3)$ \\ \hline\hline
    \end{tabular}}
\end{table}

\section{Conclusion}\label{sec:Conclusion}
In this paper, we proposed low complexity transceiver techniques for GFDM systems. The proposed transceiver techniques exploit the special structure of the modulation matrix to reduce the computational cost without incurring any performance loss penalty. In our proposed transmitter, block DFT and IDFT matrices were used to make the modulation matrix sparse and hence reduce the computational burden. We designed low complexity MF, ZF and MMSE receivers by block diagonalization of the matrices involved in demodulation. It was shown that through this block diagonalization, a substantial amount of complexity reduction in the matrix inversion and multiplication operations can be achieved. A unified receiver structure based on MF, ZF and MMSE criteria was derived. The closed form expressions for the ZF and MMSE receiver filters were also obtained. We also analyzed and compared the computational complexities of our techniques with the existing ones known so far to have the lowest complexity. We have shown that all the proposed techniques in this paper involve lower computational cost than the existing low complexity techniques \cite{GFDM_5G,GFDM_Rx}. For instance, over an order of magnitude complexity reduction can be achieved through our ZF receiver compared with the proposed technique in \cite{GFDM_Rx}. Such a substantial reduction in the amount of computations that are involved makes our proposed transceiver structures attractive for hardware implementation of the real time GFDM systems.

\appendices{
\numberwithin{equation}{section}
\section{Derivation of $\BDFT_b \bA^{\rm H}$}\label{appendix:FbAh}
The key idea in the derivation of $\BDFT_b \bA^{\rm H}$ is based on the fact that inner product of two complex exponential signals with different frequencies is zero. 
\be\label{eqn:sum}
 \sum^{N-1}_{\ell=0}{e^{j\frac{2 \pi \ell}{N}(i-k)}} = N\delta_{ik}.
\ee
From the definitions of $\BDFT_b$ and $\bA$, $\bGamma=\BDFT_b \bA^{\rm H}$ can be obtained as $\bGamma = [\bGamma_0^{\rm T},\ldots,\bGamma_{N-1}^{\rm T}]^{\rm T}$ where $\bGamma_i$'s are $M\times MN$ block matrices that can be mathematically shown as
\be \label{eqn:Gamma}
\bGamma_i = \frac{1}{\sqrt{N}}\BG^{\rm H}\sum^{N-1}_{\ell=0}\mathcal{W}^{i\ell}\BE^{\rm H}_\ell,
\ee   
where $\mathcal{W}^{i\ell}={e^{-j\frac{2 \pi i\ell}{N}}}$. Based on the definition of $\BE_\ell$ and (\ref{eqn:sum}) we have
\be\label{eqn:sum_El}
\sum^{N-1}_{\ell=0}\mathcal{W}^{i\ell}\BE^{\rm H}_\ell = N\bPsi_\kappa,
\ee
where $\kappa = {(N-i)~{\rm mod}~N}$, $\bPsi_\kappa = {\rm diag}\{[\underbrace{\bpsi_\kappa^{\rm T},\ldots,\bpsi_\kappa^{\rm T}}_{M~{\rm block~vectors}}]^{\rm T}\}$,
\begin{align*}
\bpsi_\kappa = [0,\ldots&,1,\ldots,0]^{\rm T}, \\
&\hspace{0.5mm}\uparrow \nonumber\\ 
&\kappa^{\rm {th}}{\rm ~position}\nonumber
\end{align*}
$\bpsi_\kappa$'s are $N\times 1$ vectors and $\bPsi_\kappa$ is a diagonal matrix whose main diagonal elements are made up of $M$ concatenated copies of the vector $\bpsi_\kappa$. From (\ref{eqn:sum_El}) and (\ref{eqn:sum}), $\bGamma_i$'s can be obtained as 
\be\label{eqn:Gammai}
\bGamma_i=\sqrt{N}\BG^{\rm H}\bPsi_\kappa.
\ee
Accordingly, it can be perceived that the block matrices $\bGamma_i$'s and hence the matrix $\bGamma$ are sparse. The matrix $\bGamma_i$ has only $M^2$ non-zero elements which are located on the circularly equidistant columns $\kappa,\kappa+N,\ldots,\kappa+(M-1)N$. The elements of two consecutive non-zero columns of $\bGamma_i$ are circularly shifted copies of each other. For instance, the second non-zero column of $\bGamma_i$ is a circularly shifted version of the first non-zero one by one sample. From (\ref{eqn:Gammai}), the first non-zero column of $\bGamma_i$ can be derived as $\sqrt{N}[g_{\kappa},g_{\kappa+(M-1)N},\ldots,g_{\kappa+N}]^{\rm T}$ which is the circularly folded version of the $\kappa^{\rm th}$ polyphase component of the prototype filter. One can further deduce that the matrix $\bGamma$ is a real one consisted of the prototype filter coefficients.  

\section{Closed Form Derivation of $\BD$}\label{appendix:DClosedForm}
The polyphase components of the prototype filter $\g$ can be defined as the vectors $\g_0,\g_1,\ldots,\g_{N-1}$ where $\g_i=[g_i,g_{i+N},\ldots,g_{i+(M-1)N}]^{\rm T}$. As it is shown in Appendix~\ref{appendix:FbAh}, $\bGamma = \BDFT_b \bA^{\rm H}$ is a sparse matrix with only $M$ non-zero elements in each column. The elements of $\bGamma$ can be mathematically represented as
\be\label{eqn:Gamma_ni}
[\bGamma]_{ni}=
\left\{
	\begin{array}{ll}
	\sqrt{N}[\tilde{\g}_{n^\prime}]_k, &n=\kappa M,\ldots,(\kappa+1)M-1, \\
	{}&n^\prime=i~{\rm mod}~N, \\
	{}&k=(n+M-\left\lfloor{\frac{i}{N}}\right\rfloor)~{\rm mod}~M, \\
	0, & \rm{otherwise},
	\end{array}
\right. 
\ee  
where $\tilde{\g}_{n^\prime}$ is circularly folded version of $\g_{n^\prime}$ and $\kappa = {(N-i)~{\rm mod}~N}$. From (\ref{eqn:Gamma_ni}), it can be deduced that each group of $M$ consecutive rows of $\bGamma$, i.e., $\bGamma_i$'s, whose non-zero elements are comprised of the elements of the vectors $\tilde{\g}_{n^\prime}$'s, is mutually orthogonal to the other ones. This is due to the fact that the sets of column indices of $\bGamma_i$'s with non-zero elements are mutually exclusive with respect to each other. The block-diagonal matrix $\BD$, as derived earlier in (\ref{eqn:D}), can be calculated as $\BD = \BDFT_b (\bA^{\rm H}\bA)\BDFT_b^{\rm H}$ which can be rearranged as $\BD = (\BDFT_b \bA^{\rm H})(\BDFT_b \bA^{\rm H})^{\rm H}=\bGamma\bGamma^{\rm H}$.

Due to orthogonality of $\bGamma_i$'s with respect to each other, i.e., $\bGamma_i\bGamma_j^{\rm H}=\bzero_M,~i\neq j$, it can be discerned that $\BD$ has a block-diagonal structure. Based on equation (\ref{eqn:Gamma_ni}), only equidistant columns of $\bGamma_i$'s with circular distance of $N$ are non-zero and two consecutive and non-zero columns are circularly shifted copies of each other with one sample. As a case in point, consider $\bGamma_0$ and (\ref{eqn:Gamma_ni}). Therefore, the elements $[\bGamma_0]_{00}=\sqrt{N}[\tilde{\g}_{0}]_0,~[\bGamma_0]_{(M-1)0}=\sqrt{N}[\tilde{\g}_{0}]_{(M-1)}$ and $[\bGamma_0]_{0N}=\sqrt{N}[\tilde{\g}_{0}]_{(M-1)},~[\bGamma_0]_{(M-1)N}=\sqrt{N}[\tilde{\g}_{0}]_{(M-2)}$ illustrate that the consecutive and non-zero columns of $\bGamma_0$ are circularly shifted versions of each other. Using (\ref{eqn:Gamma_ni}), one can conclude that the same property holds for the other non-zero columns of $\bGamma_0$ and all the other $\bGamma_i$'s.

The goal here is to derive a closed form for $\BD$.
\be\label{eqn:circulant}
 \BD = \bGamma\bGamma^{\rm H} = 
 \begin{bmatrix}
  \bGamma_0 \\
  \vdots  \\
  \bGamma_{N-1}
 \end{bmatrix}
 \begin{bmatrix}
  \bGamma_0^{\rm H}&
  \hdots&
  \bGamma_{N-1}^{\rm H}
 \end{bmatrix}.
\ee
$\BD$ is an $MN\times MN$ matrix comprised of $M\times M$ submatrices which are all zero except the ones located on the main diagonal, i.e., $\BD_i = \bGamma_i\bGamma_i^{\rm H}$. From (\ref{eqn:Gamma_ni}), it can be understood that the first non-zero columns of the matrices $\bGamma_i$ and $\bGamma_i^{\rm H}$ are equal to $\sqrt{N}\tilde{\g}_{\kappa}$ and $\sqrt{N}{\g}_{\kappa}$, respectively and the rest of their non-zero columns are circularly shifted version of their first non-zero column. Removing zero columns of $\bGamma_i$'s 
\be\label{eqn:Di}
\BD_i=\bGamma_i\bGamma_i^{\rm H}=\tilde{\bGamma}_i\tilde{\bGamma}_i^{\rm H},
\ee 
where $\tilde{\bGamma}_i$ and $\tilde{\bGamma}_i^{\rm H}$ are circulant matrices with the first columns equal to $\sqrt{N}\tilde{\g}_{\kappa}$ and $\sqrt{N}{\g}_{\kappa}$, respectively. Since, $\tilde{\bGamma}_i$ and $\tilde{\bGamma}_i^{\rm H}$ are real and circulant, $\BD_i$ is also a real and circulant matrix which can be obtained as 
\be\label{eqn:Dis}
\BD_i = N{\rm circ}\{{\g}_{\kappa}\textcircled{\scriptsize M}\tilde{\g}_{\kappa}\}.
\ee  

}

\bibliographystyle{IEEEtran} 

\end{document}